\input epsf
\documentstyle[12pt]{article}

\begin{document}

\begin{titlepage}
\title{A Perturbative/Variational Approach to
Quantum Lattice Hamiltonians}
\author{Jos\'e G. Esteve$^a$ and Germ\'an Sierra$^b$ \\
$^a$ Departamento de F\'{\i}sica Te\'orica,\\
Facultad de Ciencias,
Univ. de
Zaragoza. Zaragoza, Spain. \\
$^b$ Instituto de Matem\'aticas y F\'{\i}sica Fundamental,\\ CSIC,
Madrid, Spain. }

\date{}

\maketitle

\begin{abstract}
We propose a method to construct the ground state $\psi(\lambda)$
of local lattice
hamiltonians with the generic form $H_0 + \lambda H_1$,
where $\lambda$ is a coupling constant and $H_0$
is a hamiltonian with a non degenerate
ground state $\psi_0$. The method is based on the choice of an
exponential ansatz $\psi(\lambda) = {\rm exp}(U(\lambda)) \psi_0$,
which is a sort of generalized lattice version of a
Jastrow wave function. We combine perturbative and variational
techniques to get succesive approximations of the  operator
$U(\lambda)$. Perturbation theory is used  to set up
a variational method which in turn produces non perturbative results.
The computation with this kind of ansatzs leads to associate to the
original quantum mechanical problem a statistical mechanical system
defined in the same spatial dimension. In some cases
these statistical mechanical systems turn out to be integrable,
which allow us to obtain exact upper bounds to the energy.
The general ideas of our method are illustrated in the example
of the Ising model in a transverse field.

\end{abstract}

\vskip-12.0cm
%\rightline{{\bf IMAFF 10/94}}
%\rightline{{\bf UNIZAR ??/94}}
%\rightline{{\bf September 1994}}
\vskip2cm

\end{titlepage}
\def\a{\alpha}
\def\l{\lambda}
\def\i{{\rm i}}
\def\sh{ {\rm sinh}}
\def\ch{ {\rm cosh}}
\def\n{ {\bf n}}
\def\m{ {\bf m}}
\def\s{\sigma}
\def\e{ {\rm exp} }
\def\V{ {V}_I }
\def\p{ \psi}
\def\L{{ \cal{L}} }
\def\t{ {\rm tanh}}
\def\en{ {\rm e}}
\def\b{ \beta}
\def\var{ {\rm var}}
\def\per{ {\rm per}}

\section{Introduction}

In Statistical Mechanics and Quantum Field Theory there is a
large class of two-dimen\-sional models
whose basic magnitudes
can be computed exactly\cite{Ba}. The mathematics and physics of these
integrable models are very rich, constituing areas
of intensive investigation.
There are in contrast very few integrable models
in higher dimensions\cite{Za,BaBa}, where almost all the models
which are interesting from a physical point of view
are believed to be non integrable. These facts represent a
major conceptual and technical
gap between the 2d and the higher dimensional worlds,
which would be desirable to fill in.

The lack of exact techniques has always motivated the construction
of approximative methods which very often give good quantitative
and qualitative results.
These methods can be roughly classified according to four main
categories,
namely: perturbative, variational, numerical and
renormalization group, every one having
advantages and limitations. The perturbative methods can be developed
in a systematic way
but are restricted to small values of the coupling constants.
The variational ones give in general non perturbative
results but rely on aprioristic or
intuitive  conjectures.
The numerical approaches are in general unbiased but
limited to small lattices. Finally, the
RG techniques  when combined with one of the previous ones
are very powerful for the study of critical phenomena.

In this paper we shall restrict ourselves to the first two
methods. Our aim is to combine perturbative and variational
techniques choosing the best of each and by passing
some of their limitations.
The problem we cope with is the construction
of the ground state of a local
lattice  hamiltonian of the form $H_0 + \l H_1$,
where $\l$ is a coupling constant.
Our  strategy  is to use perturbation
theory to set up a variational
approach to the computation of the ground state wave function.
The starting point is the choice of an exponential ansatz
for the exact ground state of
the hamiltonian
$H_0 + \l H_1$.
The peculiarity of this
ansatz is that,  under certain assumptions about the ground
state of the unperturbed hamiltonian $H_0$ ,
the whole hierarchy of perturbative equations can be
solved systematically in terms of a collection of
irreducible local operators $ \V $,
which are  characterized
by the order $\nu_I$
at which the operator $\V$ first appears in perturbation
theory.
Each operator $\V$ comes in the
ansatz multiplied by a weight
$\a_I(\l)= \l^{\nu_I}+$ higher powers.
The data  $\{ \V, \a_I(\l) \}$ contain all the
information required to built up the ground state.
For small lattices it is easy to construct
this set and therefore obtain the exact ground state,
but for large lattices we have to resort to some kind
of approximation. The natural thing to do is to truncate
the ansatz considering only those operators
$\V$ with a level equal or less that a given number $\nu$,
which is the order in perturbation theory where one is working at.
The problem is then to find the weights $\a_I(\l)$
for $\nu_I \leq \nu$.
At this point one can follow two different paths: i)
use the truncated ansatz as a trial state in a standard
variational fashion or ii) take the truncated ansatz
as the exact ground state of a hamiltonian which should
be sufficiently close to the original one.
In this paper we shall develop the approach i) and leave
the approach ii) for another publication.
We have so far outlined the main ideas and techniques
which will be exemplified in the case of the Ising
model in a transverse magnetic field (ITF), but it should be
clear from the exposition that they apply to more complicated
models as for example the chiral Potts model, the XXZ- model, etc.

Trial wave functions with an
exponential or, more generally,  product structure
are known as Jastrow wave functions and has been applied
to a wide variety of problems in different
areas as for example,
liquid $^4$He \cite{Ma}, nuclear
matter \cite{Ja}, solid state physics \cite{Sut,KKLZ}
and more recently
in the fractional quantum Hall effect
\cite{La}.
In quantum mechanics a Jastrow wave function is simply the
exponential of the potential energy, $\psi(x) =\e(-\a V(x))$
and becomes exact for the simple harmonic oscillator. In
a many body problem one replaces $V(x)$ in the exponential
by $\sum_{i,j}
f(x_i - x_j)$ where $f$ is some two body effective potential.
In this paper we shall consider discrete versions of generalized
Jastrow states in contrast of the continuous versions described
above.
To our knowledge there is not in the literature
a general and systematic study
of this type of wave functions except for some interesting observations
and discussions confined to particular models \cite{ArGi}.
Among these observations is the fact
that the norm of the Jastrow wave functions coincides, in certain cases,
with  the partition function of an associated classical statistical
mechanical model defined in the same dimension, with the variational
parameter playing the role of inverse temperature.
Indeed, in the example of the ITF
we find that the norm of the exponential ansatz
is given
by the partition function of a classical Ising model with couplings
among the spins dictated by the familly of operators
$ \V $ and
as many coupling constants as weights $\a_I$.
Working with these generalized Jastrow wave functions
requires the knowledge of thermodynamic quantities such as
internal energies, susceptibilities, etc.
The calculation of these quantities
is in general a difficult task and this imposes limitations to
the variational approach i) mentioned above.
However in certain occasions the computation can be carried out exactly.
This happens when the associated statistical
system which underlies
a particular ansatz is integrable. If the quantum hamiltonian
is one dimensional then the statistical system is also one dimensional
and the corresponding partition function can be easily computed.
Things are more interesting in two dimensions where,
as we said in the introduction,
most of the hamiltonians are not exactly solvable.
The approximation method that we propose bring us to statistical
systems in the same dimension, and therefore in
two dimensions we may some times make contact with 2d integrable
models. This opens the possibility of using the huge amount
of information known in integrable 2d systems in the study of
2d quantum systems.

The paper is organized as follows. In section 2 we introduce the
P/V method. In section 3 we apply the general ideas
presented in section 2 to the Ising model in a transverse
magnetic field, making special emphasis in the 1d and 2d cases.
In section 4 we present our conclusions and future prospects.
All the technical details and computations are included
in four appendices.

\section{The Perturbative/Variational Method}

Let $H(\l) = H_0 + \l H_1$ be a local lattice hamiltonian
where $H_0$ is a Hamiltonian whose ground state $\psi_0$
is non degenerate. We suppose that $\psi_0$ and its energy
$E_0$ are known exactly. If $\l$ is sufficiently small one can
use standard perturbation theory to expand the ground state
wave function $\p(\l)$ and energy $E(\l)$
of  $H(\l) = H_0 + \l H_1$ as follows,

\begin{eqnarray}
& \p(\l) = \sum_{n=0}^{\infty} \l^n \psi_{n} &
\label{1} \\
&E(\l) = \sum_{n=0}^{\infty} \l^n E^{(n)}&
\label{2}
\end{eqnarray}

\noindent
where $\{ \p_{n} \}$ and $\{ E^{(n)} \}$ are subjected to
satisfy a set of equations linking quantities of the
same order \cite{GaPa}.
In order to write this set of equations we shall
assume that $\p(\l)$ can be reached from $\p_0$
with the action of an operator of exponential type,

\begin{eqnarray}
&\p(\l) = e^{U(\l)} \; \p_0 & \nonumber \\
& U(\l) =  \sum^{\infty}_{n=1} \l^n  U_n &
\label{3}
\end{eqnarray}

The relation between the perturbative expansion (\ref{1})
and the operators $U_n$ is given by a non-abelian version
of the Schur polynomials,

\begin{eqnarray}
&\p_1 =  U_1 \p_0 & \nonumber \\
&\p_2 =  \left( \frac{1}{2} U^2_1 + U_2 \right) \p_0 & \label{4} \\
&\vdots  & \nonumber \\
&\p_n =  \sum_{p=1}^n \frac{1}{p!} \sum_{n_1+ \cdots +n_p=n} \;
U_{n_1} \cdots U_{n_p} \; \p_0 \nonumber
\end{eqnarray}

If  the operators  $U_n$ commute among themselves
then the above expressions coincide with the Schur polynomials
$S_1, S_2, \cdots, S_n$
as functions of the variables $U_n$.
The equations satisfied by $\p_n$ and $E^{(n)}$ translate now into
a set of equations for $U_n$ and $E^{(n)}$ which we pass to derive.

Introducing (\ref{3}) into the eigenvalue equation,

\begin{equation}
H(\l) \; \p(\l) = E(\l) \; \p (\l)
\label{5}
\end{equation}

\noindent
and multiplying on the left hand side by the inverse of
the exponential one gets,

\begin{equation}
\e \left( -
\sum^{\infty}_{n=1} \l^n U_n \right) \;
H(\l) \;
\e \left(
\sum^{\infty}_{n=1} \l^n U_n \right)
\p_0 = E(\l) \; \p_0
\label{6}
\end{equation}

\noindent
To simplify
(\ref{6}) we use the identity \cite{Wi},

\begin{equation}
e^{-A} \; B \; e^{A} = e^{\L_A} (B) = \sum^{\infty}_{n=0}
\frac{1}{n!} \; \L^n_A (B)
\label{7}
\end{equation}

\noindent
where

\begin{equation}
\L_A(B) = [ B,A ] \; ,\;\; \L^2_A = [ [B,A],A], \dots
\label{8}
\end{equation}

\noindent
which yields,

\begin{equation}
\e \left( \sum^{\infty}_{n=1} \l^n \L_{U_n} \right)\;
H(\l)\;\; \p_0 = E(\l) \; \p_0
\label{9}
\end{equation}

The equations for the operators $U_n$ are finally
obtained expanding
the exponential in (\ref{9}) and collecting all the
terms of the same power in $\l$ ,

\begin{eqnarray}
& H_0 \; \p_0 = E_0  \;\p_0 & \nonumber \\
&\left( [H_0 , U_1 ] + H_1 \right) \p_0 = E^{(1)} \;\p_0 &
\nonumber \\
&\left( [H_0, U_2] +
\frac{1}{2} [ [H_0 , U_1 ], U_1]  +
[H_1, U_1]  \right) \p_0 = E^{(2)} \;\p_0 & \label{10} \\
& \vdots & \nonumber \\
& \left[ \sum^n_{p=1}  \sum_{n_1+ \cdots+ n_p=n} \; \frac{1}{p!} \;
\L_{U_{n_1} } \; \cdots \L_{U_{n_p}} (H_0) \right.
&\;\;\;\;\;\;\;\;{\rm{ for}\; n \geq 2}  \nonumber \\
& \left.+  \sum^{n-1}_{p=1}  \sum_{n_1+ \cdots+ n_p=n-1} \;
\frac{1}{p!} \;
\L_{U_{n_1} } \; \cdots \L_{U_{n_p}} (H_1) \right] \; \p_0
= E^{(n)} \; \p_0  & \nonumber
\end{eqnarray}

The solution of these equations is by no means unique.
In particular the hermiticity properties of $U(\l)$
are not fixed a priori. There are two main
choices that can be made
in terms of
hermitean or antihermitean operators.
If $U(\l)$ is choosen to be
antihermitean, then eq.(\ref{3})
becomes a unitary transformation which preserves obviously
the norm of the
state, $ i.e. <\p(\l)|\p(\l)> = <\p_0|\p_0>=1$.
Moreover if
that $H_1$ has an off diagonal form, then $E^{(1)}$
vanishes. When this happens it is customary to perform a
unitary transformation,

\begin{equation}
H_0 + \l H_1 \rightarrow e^{- \i \l S} (H_0 + \l H_1)
e^{\i \l S} = H_{{\rm eff}}
\label{11}
\end{equation}

\noindent
where $S=S^{\dagger}$ is obliged to satisfy,

\begin{equation}
\i [ H_0 , S ] + H_1 = 0
\label{12}
\end{equation}

\noindent
in order to cancel the
terms proportial to $\l$ in $H_{\rm eff}$.
This equation coincides with the first order
equation in (\ref{10}) but expressed in a
matrix form, i.e. the dependence on the state
$\p_0$ has been dropped.
Finally the "effective hamiltonian" $H_{\rm eff}$ has the expansion,

\begin{equation}
H_{\rm eff} = H_0 + \frac{\i \l^2}{2} [H_1, S] + O(\l^3)
\label{13}
\end{equation}

These kind of unitary transformations
are extensively used in Statistical
Mechanics and
Condensed Matter
Physics\cite{BEMS}, and
have also been considered from a more mathematical point
of view \cite{Ka}.

In this paper we are interested in the study
of those solutions to equations
(\ref{10}) which are hermitean.
An immediate consequence of this choice is that
the norm of the state $\p(\l)$ is no longer the unity
and in fact it will be an important quantity to compute,

\begin{equation}
Z(\l) = < \p(\l)| \p(\l)> = <\p_0|e^{2 U(\l)} |\p_0>
\label{14}
\end{equation}

A peculiarity of equations (\ref{10})
is that all the operators $U_n$ appear always
involved in nested commutators.
Let us suppose  that $H_0$ and $H_1$ are local
operators defined
on a lattice, i.e. operators involving a finite number
of lattice variables located within a neighbourhood,
whose size is independent of the size of the
whole lattice. This kind of operators go, in the thermodynamic
limit, to local operators in the continuum.
We shall also assume that the number of degrees of freedom per site
is finite.
We want to investigate
under which conditions
the solution of the equations (\ref{10}) can be given
by a family  of local operators.
That this kind of solutions
may exist is suggested by  the property mentioned
above about the nested structure
of equations (\ref{10}).
The reason is that the commutator of local operators is again
a local operator and therefore each equation (\ref{10})
reduces to a sum  of local operators acting on the unperturbed
ground state.
The other ingredient we shall require to achieve
a local solution is an unperturbed ground state $\p_0$
with an ordered structure, which can be
ferromagnetic, antiferromagnetic, or of
some more general type. This is sufficient to guarantee that
the operators $H_0, H_1 $ and $U_n$
should produce local disturbances acting on $\p_0$.
There are many physical systems where these conditions are
fullfilled. The reason for  requiring an ordered ground state is
that in this case equations (\ref{10}) become
a system of linear
equations with a number of
unknowns depending on the dimension of the lattice but not
on its size. It will also depend on the order of perturbation
theory and the number of lattice variables at each point.
The aim of the above considerations
is  to make plausible the existence of
solutions of the perturbative equations in terms of local
operators $U_n$, but of course it does not constitute a proof.
Later on we shall go through particular
examples to show that this kind of solutions do indeed exist.
In the case of the Ising
model , and we suspect it is a general feature ,
every operator $U_n$ is given by a
sum of irreducible local operators
which we denote by $\V$,

\begin{equation}
U_n = \sum_I \; p_{I,n} \; V_I
\label{15}
\end{equation}

\noindent
where the coefficients $p_{I,n}$ are
obtained by solving equations (\ref{10}).
and where each operator $\V$
is a product of Pauli matrices located at different
points of the lattice.
The sum over $I$ in (\ref{15}) is finite
and contains in general operators that have already appeared
in the solution of equations of lower order. It is convenient
to define the level $\nu_I$ of  $\V$ as the order
in perturbation theory at which this
operator appears for the first time in the
solution of equations (\ref{10}),

\begin{equation}
\nu_I = {\rm minimum}\; \{ n \; / \; p_{I,n} \neq 0 \}
\label{16}
\end{equation}

Introducing (\ref{15}) into (\ref{3}) and interchanging the order
of the sums $ \sum_n \sum_I = \sum_I \sum_n$,
we rewrite the exponential ansatz as follows,

\begin{equation}
\p(\l) = \e \left( \frac{1}{2} \; \sum_I \; \a_I (\l) \V \right) \p_0
\label{17}
\end{equation}

\noindent
where  $\a_I(\l)$ is the "weight" of $\V$ in the ground state,

\begin{equation}
\a_I(\l) = 2 \;\sum^{\infty}_{n=\nu_I} \; p_{I,n} \l^n
\label{18}
\end{equation}

The factor $1/2$ in eq.(\ref{17}) has been introduced for later
convenience.
Definition (\ref{18}) implies that $\a_I(\l)$ has a Taylor expansion
starting at $\l^{\nu_I}$, and containing non vanishing terms of
higher powers whenever $\V$ appears in the solution of eqs.(\ref{10}).
The knowledge of a single function $\a_I(\l)$
will in general require
solving the entire hierarchy (\ref{10})!.
At this stage one may wonder whether there are other alternatives
to compute the functions
$\a_I$ rather than solving the whole hierarchy (\ref{10}).
To handle this problem
it is illustrative
to compare equations (\ref{3}) and (\ref{17}). They
both give two different perturbative
approximations to the ground state.
While (\ref{3}) is an expansion
in the coupling constant $\l$,
eq.(\ref{17}) is a kind of cluster expansion.
In a sense
they are conjugated perturbative expansions.
Exploiting this analogy we shall propose
an alternative way for computing
$\a_I(\l)$. In standard perturbation theory one truncates
the state excluding corrections beyond a given power $\l^n$.
The analog of this for (\ref{17}) will be to
truncate the sum over $I$
to contain only those operators $\V$ with a level $\nu_I$ less or
equal than a given order $\nu$,

\begin{equation}
\p^{(\nu)}(\l) = \e \left( \frac{1}{2} \; \sum_{I, \nu_I \leq \nu}
\a_I^{(\nu)}( \l) \; \V \right) \; \p_0
\label{19}
\end{equation}

In eq.(\ref{19}) we leave open the possibility that the
weights $\a_I^{(\nu)}$ depend on the order of the approximation
$\nu$. The problem now is how to determine
$\a_I^{(\nu)}$. In this paper we shall use
a variational
method which consist in taking (\ref{19}) as a trial wave function
for the ground state of the hamiltonian $H_0 +\l H_1$.
$\a_I^{(\nu)}$ are of course the variational parameters.
Minimization
of the energy of this state will give the weights $\a_I^{(\nu)}$
as functions of $\l$. It is clear that this minimization procedure
should agree with the result
obtained in
perturbation theory to the same degree
of accuracy, i.e.

\begin{equation}
\a_I^{(\nu)}(\l) - \a_I(\l) = O(\l^{\nu+1}) ,\; \;\;
\forall I \;\; / \;\;
\nu_I \leq \nu
\label{20}
\end{equation}

On the other hand,
it is well known that if one takes the perturbative solution
to an order
$\nu$ as a trial wave function then the variational energy
agrees with the perturbative expansion to order $2 \nu +1$ \cite{MoFe},

\begin{equation}
E^{(var,\nu)}(\l) - E(\l) = O( \l^{2  \nu +2})
\label{21}
\end{equation}

This fact already illustrates at
the perturbative level the usefulness of
combining both perturbative and variational
techniques\footnote{Recall that in  perturbation theory
the knowledge of the $n^{th}$ order correction
of the wave function allows the
computation of the energy
to order $n+1$ according to the formula
$E^{(n+1)} = <\psi_0|H_1|\psi^{(n)}>$ .}.
Next we consider the non-perturbative
aspects of (\ref{19}).
The question is whether this ansatz can be extended to large
values of $\l$ remaining close to the exact ground state.
In particular in the limit $\l \rightarrow \infty$ we wish the
state $\p^{(\nu)}$ to flow to the ground state manifold of the
Hamiltonian $H_1$, i.e.

\begin{equation}
\lim_{\l \rightarrow \infty} H_1 \p^{(\nu)}(\l) = E_1 \;
\lim_{\l \rightarrow \infty} \p^{(\nu)}(\l)
\label{22}
\end{equation}

\noindent
where $E_1$ is the ground state energy of $H_1$.
If this equation holds then $\p^{(\nu)}(\l)$ will interpolate
in a continuous way the ground state of $H_0$ and those of
$H_1$, giving certain validity
to the results obtained for  the intermediate values of
the coupling constant $\l$, where
a weak or a strong coupling expansion are dubious.
A sufficient set of conditions for eq.(\ref{22}) to hold is the existence,
within the family of operators expanded by $\{\V \}$,
of an hermitean operator
$\sum_{I, \nu_I \leq \nu_0}
a^{(\nu_0)}_I \; \V $ having a common eigenvector, say $\phi_0$,
with the Hamiltonian $H_1$, and such that $\phi_0$ and $\psi_0$
have non-vanishing overlap,

\begin{equation}
\begin{array}{ll}
i) &  \sum_{I, \nu_I \leq \nu_0}
a^{(\nu_0)}_I  \; \; \V \; {\rm is } \; {\rm  hermitean}  \\
ii) &   \phi_0 \;\;  {\rm is} \; {\rm a} \; {\rm ground} \;
{\rm state} \; {\rm of} \; H_1 \; {\rm and} \;
\sum_{I, \nu_I \leq \nu_0}
a_I^{(\nu_0)} \; \V \\
iii) & <\p_0|\phi_0> \neq 0 \end{array}
\label{23}
\end{equation}

\noindent

Under these conditions it is easy to see that the ansatz
$\p^{(\nu)}(\l)$ for $\nu \geq \nu_0$
flows to  $\phi_0$
in the limit $\l \rightarrow \infty$.
The interest of conditions (\ref{23}) is that they strongly
restrict the possible solutions of the perturbative equations
(\ref{10}) (see appendix C for concrete examples).

\section{The Ising Model in a transverse field}

\subsection{General Considerations}

In this section we shall study in detail the Ising
model in a transverse field.
Let us introduce our notations and some generalities
valid in any dimension.

The model is defined
in a hypercubic lattice in d dimensions and $L$ sites
with a Hamiltonian  given by \cite{dG,Pe1,St},

\begin{equation}
H_d=- \l \; \sum_{\n}  \; \s^X_{\n} - \sum_{\n, {\bf \mu}} \;
\s^Z_{\n} \; \; \s^Z_{\n+\mu}
\label{24}
\end{equation}

\noindent
where  $\s^X_{\n}$ and $\s^Z_{\n}$ are Pauli matrices acting
at the site $\n$,  with
$\n =(n_1,\dots, n_d)$,  $n_a=1,\dots, L^{1/d}$  and
${\bf \mu}$ is any of the lattice vectors $\mu_1= (1,0,\dots,0) , \dots,
\mu_d= (0,0,\dots,1)$. The cases $d=1$ and $2$ correspond to a linear chain
and a rectangular planar lattice respectively.
We shall suppose periodic boundary conditions.

The phase diagram of this Hamiltonian as a function of $\l$ and
the temperature is well known \cite{ElWo} . We shall restrict ourselves to
zero temperature, whose analysis is interesting because it shows
a simple but non trivial example of quantum critical phenomena
\cite{Pe2,PeEl}. In this case there is a critical transverse field
$\l_c$ below which there are two ferromagnetic ground states
characterized by the non vanishing of $<\s^Z_{\n}>$.  For $ \l >\l_c$ the
ground state is disordered and $<\s^Z_{\n}> = 0$ but
$<\s^X_{\n}> \neq 0$. Near $\l_c$ the critical behaviour of the
$1d$ Ising model with transverse field as a function of $\l$
at $T=0$ is the same as the critical behaviour of the classical
Ising model
in two dimensions as a
function of $T$ \cite{Pe2}.
This can be seen from the fact
that the Ising model hamiltonian in 1d is related to the
transfer matrix of the 2d Statistical Ising model.
The analogy between the quantum critical behaviour of the
d-dimensional ITF and the statistical
critical behaviour of the Ising model in d+1 dimensions for d=1
is believed to hold for $d \geq 1$ \cite{Pe2,EPW,Suz}.
This fact is not peculiarity of the ITF but a property which
holds more generally \cite{FaSu,Ko}.

In $1d$ the model has been solved exactly
by many different methods \cite{On,SML,Pe2}. The ground
state energy of a periodic chain of $L$ sites is ,

\begin{equation}
E^{(1d)}_0 = -  \; \sum^{L/2-1}_{n= -L/2}
\sqrt{ 1 + \l^2 + 2 \l \cos \left[ \frac{(2n +1) \pi}{L} \right]}
\label{25}
\end{equation}

\noindent
which gives the following energy density in the thermodynamic limit,

\begin{equation}
\en^{(1d)}_0 =\lim_{L \rightarrow \infty}
\frac{E^{(1d)}_0}{L}=
- (1 + \l) \; F\left(-\frac{1}{2}, \frac{1}{2}; 1; \frac{4 \l}{(1 +
\l)^2} \right)
\label{26}
\end{equation}

Eqs.(\ref{25}) and
(\ref{26}) exhibit the Krammers-Wannier duality,

\begin{equation}
\en^{(1d)}_0(\l)= \l\; \en^{(1d)}_0(\frac{1}{\l})
\label{27}
\end{equation}

\noindent
in agreement with the fact that $\l_c=1$ is the selfdual
critical transverse field. The density energy at the
critical field is,

\begin{equation}
\en^{(1d)}_0 = - \frac{4}{\pi}
\label{28}
\end{equation}

For  dimensions higher than one the exact
solution is not known
and only approximations are available \cite{PeEl}.
In the next subsections we shall obtain
variational upper bounds for the exact energy. It is also
possible to obtain a lower bound for the ground state energy
of the Hamiltonian (\ref{24}),

\begin{equation}
\en^{(d)}_0(\l) \geq  d \; \en^{(1d)}_0 \left(
\frac{\l}{d} \right)
\label{28b}
\end{equation}

There are two extreme cases where one can find the
exact ground state of the Hamiltonian (\ref{24}).
At $\l =0$ there are two ferromagnetic ground states
corresponding to $<\s^Z> = \pm 1$ given by,

\begin{equation}
|  \uparrow > =  \otimes_{\n} \;\;|\uparrow >_{\n} \;\;, \;\;
|  \downarrow > = \otimes_{\n} \;\; |\downarrow >_{\n}
\label{29}
\end{equation}

\noindent
where $|\uparrow >_{\n} \;, |\downarrow >_{\n}$ denote the
eigenstates of $\s^Z_{\n}$ with eigenvalues $1, -1$ respectively.
The density energy of these states is,

\begin{equation}
\en_0(\l=0) = -  \; d
\label{30}
\end{equation}

At $\l = + \infty$ the ground state is the non  degenerate
disordered state ( $< \s^Z> =0$ ),

\begin{equation}
| 0 > = \otimes_{\n} |0>_{\n} = \otimes_{\n} \;\;  \frac{1}{\sqrt{2}}
( |\uparrow >_{\n} + |\downarrow >_{\n} )
\label{31}
\end{equation}

\noindent
whose energy is given in the limit $\l \rightarrow \infty $ by,

\begin{equation}
\en_0(\l \rightarrow + \infty) \simeq - \; \l
\label{32}
\end{equation}

The hamiltonian (\ref{24}) commutes with
the spin rotation operator,

\begin{equation}
R = \prod_{\n} \;\; \s^X_{\n}
\label{33}
\end{equation}

\noindent
and therefore its spectrum can
be split into an even $(R=1)$  and odd $(R=-1)$ sectors.
The ground state (\ref{31}) of the disorder phase
belongs to the even sector
whereas in the ordered phase there is a ground state belonging to
each of the two sectors.

We shall study the ordered and disordered phases independently.
In the next two subsections we present the results
of the computation of the energy and the variational parameters
using the variational method. The details
of their derivation will be presented in the appendices.
The formulae for the energy and the weights $\a_I$ are labelled
by the type of approximation used, its order and the spatial
dimension of the lattice, namely,
\en(app, order, d), $\a_I$(app,order,d),
where
app=  per (perturbative), var (variational).

\subsection{Ordered Phase}

In this phase the ground state is degenerate and therefore
we could not in principle apply the techniques of section 2.
However in practice this is not an obstacle since for large lattices
the state $|\uparrow>$ does not mix with the state
$|\downarrow>$ at finite orders in perturbation theory and
one can effectively apply non degenerate perturbation theory,
say to the state $| \uparrow >$ \cite{PeEl}.
In the ordered phase $\l$ is a small parameter and therefore
the hamiltonian (\ref{24}) split as follows,

\begin{equation}
\begin{array}{cl}
H = & H_0 + \l H_1 \\
H_0 =& - \sum_{\n} \sum_{\bf \mu} \; \s^Z_{\n} \; \s^Z_{\n +
\mu} \\
H_1 =& - \sum_{\n} \s^X_{\n}
\end{array}
\label{34}
\end{equation}

\noindent
The unperturbed ground state is taken to be,

\begin{equation}
\p_0 =  |\uparrow > = \prod_{\n} |\uparrow>_{\n}
\label{35}
\end{equation}

\noindent
A solution of the perturbative equations (\ref{10})
which satisfy also conditions (\ref{23})
is given by,

\begin{equation}
\begin{array}{cl}
U_1 =& \frac{1}{4 d} \sum_{\n} \s^X_{\n}  \\
\end{array}
\label{36}
\end{equation}

\begin{equation}
\begin{array}{cl}
U_2 =& \frac{1}{16 d^2 (2d-1)} \sum_{\n, \mu}
\s^X_{\n} \s^X_{\n + \mu}  \\
\end{array}
\label{37}
\end{equation}

In appendix C we consider another solutions
to equations (\ref{10}),
which do not satisfy (\ref{23}), and discuss their
significance.

The perturbative energy to third order
in $\l$ which can be derived from this solution is,

\begin{eqnarray}
\en(\per, n=3, d ) = -
\left( d + \frac{ \l^2}{ 4 d}   \right)
\label{38}
\end{eqnarray}

Notice that odd powers in $\l$ do not enter into the energy.

We shall now apply the P/V method
corresponding to  $\nu= 1$ and $2$.
{}From eq.(\ref{36}) the ansantz for $\nu=1$ is given by,

\begin{equation}
\begin{array}{rl}
\p^{(1)} = & \e  \left( \frac{h}{2} \sum_{\n} \s^X_{\n} \right)
|\uparrow> \\
= &  \prod_{\n} \left( \ch \frac{h}{2} \; |\uparrow>_{\n} +
\sh \frac{h}{2} \; |\downarrow>_{\n} \right)
\end{array}
\label{39}
\end{equation}

\noindent
which shows that $\p^{(1)}$ is a mean field state parametrized
in terms of hyperbolic functions of the variational
parameter $h$.
This mean field state is usually written
in terms of an angle $\theta$,
which describes the deviation of the up pointing spin under
the action of the transverse field \cite{dG}.
These two kinds of parametrizations
can be explained from
the hermiticity properties of $U_1$, i.e.
$h$(non-compact variable) $\leftrightarrow$ $U_1$(hermitean),
$\theta$(compact variable) $\leftrightarrow$ $U_1$(antihermitean)
(see appendix C).

The minimization of the energy of (\ref{39}) is achieved for,

\begin{equation}
\en(\var ,\nu=1,d ) =
\left\{  \begin{array}{ll}
- (d + \frac{\l^2}{4 d}) & \l \leq 2 d \\
-\l & \l \geq 2 d
\end{array} \right.
\label{40}
\end{equation}

\noindent
For $\l \leq 2d $ the variational energy (\ref{40})
coincides with the perturbative
result (\ref{38}), in agreement with condition (\ref{21}).
For $\l \geq 2d$ this energy coincides with the strong coupling limit
(\ref{32}), in agreement now with eq.(\ref{22}).

The mean field state (\ref{39})
illustrates in the simplest possible way the interpolating properties
of the variational states constructed
out from solutions of  the perturbative equations
(\ref{10}) fulfilling conditions
(\ref{23}). Later on we shall see more
examples of this.

The dependence of the variational
parameter $h$ on $\l$ is given by,

\begin{equation}
h(\var ,\nu=1,d) =
\left\{ \begin{array}{cl}
{\rm arctanh} \left(\frac{\l}{2 d} \right) & \l \leq 2d \\
\infty & \l \geq 2d
\end{array} \right.
\label{41}
\end{equation}

The value $h= \infty$ for $\l \geq 2d$ implies that $\p^{(1)}$
can be identified with the state $|0>$, which is precisely the eigenstate
of the Hamiltonian $H_1$.

The magnetization of (\ref{39}) is,

\begin{equation}
< \s^Z > = \left\{ \begin{array}{cl}
\left[ 1 - \left( \frac{\l}{2 d} \right)^2
\right]^{1/2} & \l \leq 2d \\
0 & \l \geq 2d \end{array} \right.
\label{42}
\end{equation}

\noindent
which shows that $\l_c = 2d$ and the exponent $\beta$,
defined by $< \s^Z > \sim (\l_c - \l)^{\beta} $, takes the
mean field value $\beta = 1/2$.
The exact result in 1d is given by,

\begin{equation}
< \s^Z > = \left\{ \begin{array}{cl}
\left( 1 - \l^2 \right)^{1/8} & \l \leq 1 \\
0 & \l \geq 1 \end{array} \right.
\label{42b}
\end{equation}

To go beyond the mean field approximation we shall consider the
second order ansatz  $\nu =2$, which according to (\ref{37}) reads

\begin{equation}
\p^{(2)} = \e  \left( \frac{h}{2} \sum_{\n} \s^X_{\n}
+ \frac{\a}{2}  \sum_{\n, \mu} \s^X_{\n} \s^X_{\n +\mu} \right)
|\uparrow>
\label{43}
\end{equation}

The term in the exponential of (\ref{43}) proportional to
$\alpha$ introduces in $\p^{(2)}$ correlations which were
absent in $\p^{(1)}$. Hence $\p^{(2)}$ is no longer a mean field state.
The norm of $\p^{(2)}$ coincides with the partition function
of a  d-dimensional
classical Ising model with inverse temperature $\a$ and magnetic
field $h$. In $1d$ the computations can de done exactly (see
appendix A). The results for the energy and magnetization are
shown in figure 1 and 2. The variational estimation of the
energy for $\nu=2$ considerably improves the mean field value result.
The magnetization is also improved, but the singular mean
field behaviour is lost obtaining instead
a smooth curve (see table 1).
This implies that in order to obtain an estimation
of the exponent $\beta$ we should resort to another
techniques as for example $Pad\acute{e}$ approximants.
The study of these questions,
though interesting,
is beyond the scope of this paper.

In $d \geq 1$ the norm of the state (\ref{43}) cannot be computed
exactly hence we have to use some kind of approximation. For small values
of $\l$ we expect both $h$ and $\alpha$ to be also small, in which
case this norm can be computed through a "high temperature"
expansion. Doing this computation we obtain,

\begin{equation}
\en(\var ,\nu=2,d ) =
- \left( d + \frac{\l^2}{4 d} +\frac{\l^4}{64 d^3(2d -1)}
+ O(\l^6) \right)
\label{43bis}
\end{equation}

\noindent
The fourth order term in $\l$
agrees with the perturbative result
in accordance with (\ref{21}).

For large values of $\l$ we expect that $h$, $\a$ or
both become also very large in which case the norm
of the variational state can be
computed by means of a "low temperature" expansion.
Of course if the model is integrable we can compute in the
whole range of couplings including the intermediate ones
, obtaining rigorous upper bounds for the exact energy.

\subsection{Disordered Phase}

In this phase the ground state is
non degenerate and hence the techniques
explained in section 2 could be automatically applied .
The perturbative parameter in the disordered  phase is $\frac{1}{\l}$,
which leads us to write the Hamiltonian (\ref{24}) as,

\begin{equation}
\begin{array}{rl}
H = & \l \left( H_0 + \frac{1}{\l} H_1 \right) \\
H_0 = & - \sum_{\n} \; \s^X_n \\
H_1 = & - \sum_{\n, \mu} \; \s^Z_{\n} \; \s^Z_{\n + \mu} \\
\end{array}
\label{44}
\end{equation}

The unperturbed ground state is,

\begin{equation}
\p_0= |0> = \prod_{\n} |0>_{\n}
\label{45}
\end{equation}

A solution of the equations (\ref{10}) and (\ref{23}) reads,

\begin{equation}
\begin{array}{rl}
U_1 = & \frac{1}{4}  \sum_{\n, \mu} \; \s^Z_{\n} \; \s^Z_{\n +
\mu}
\end{array}
\label{46}
\end{equation}

\begin{equation}
\begin{array}{rl}
U_2 = & \frac{1}{16}  \left(
\sum_{\n, \mu} \; \s^Z_{\n} \; \s^Z_{\n + 2 \mu} +
\sum_{\n, \mu_1  \neq \mu_2} \; \s^Z_{\n} ( \s^Z_{\n + \mu_1 + \mu_2}
+ \s^Z_{\n + \mu_1 - \mu_2} ) \right)
\end{array}
\label{47}
\end{equation}

The perturbative energy to third order is,

\begin{equation}
\en (\per , n=3, d ) = -  (
\l + \frac{d}{ 4 \l}  )
\label{48}
\end{equation}

In this phase we shall only study the first order ansatz ($\nu =1$)
which according to (\ref{46}) becomes,

\begin{equation}
\begin{array}{cl}
\p^{(1)} =& \e \left( \frac{\a}{2}  \sum_{\n,\mu}
\s^Z_{\n} \s^Z_{\n+ \mu} \right)
|0>
\end{array}
\label{49}
\end{equation}

\noindent
The norm of $\p^{(1)}$ coincides with the partition function of the
d-dimensional classical Ising model with inverse temperature $\a$.
This implies that computations with (\ref{49}) can only be done
exactly in $d=1$ and $2$.

The 1d case is very easy to handle ( see appendix B). The energy of
(\ref{49}) is after minimization,

\begin{equation}
\en(\var , \nu =1, d=1) = \left\{
\begin{array}{cl}
- \left( \l + \frac{1}{4 \l} \right) & \l \geq 1/2 \\
-1 & \l \leq 1/2  \end{array} \right.
\label{50}
\end{equation}

Compairing this result with the corresponding one in the ordered
phase (\ref{40}), we observe that they satisfy the duality relation
(\ref{27}), which means that we
have incorporated the KW duality in our ansatzs.
We have now two kinds of variational results at order $\nu =1$,
which can be combined in order to optimize the upper
bound to the energy,

\begin{eqnarray}
\en(\var , \nu =1, d=1) =& \\{\rm Minimum}& \left(
\en^{({\rm order})}(\var , \nu =1, d=1),
\en^{({\rm disorder})} (\var , \nu =1, d=1) \right)  \nonumber \\
&= \left\{
\begin{array}{cl}
- \left( 1 + \frac{\l^2}{4} \right) & \l \leq 1 \\
- \left( \l + \frac{1}{4 \l} \right) & \l \geq 1 \\
\end{array}  \right.
\label{51}
\end{eqnarray}

Let us define $\l^{(\nu)}_{OD}$ as the crossing point between
the ordered and disordered variational energies at order
$\nu$. We expect that  $\l^{(\nu)}_{OD}$
will approach $\l_c$ as $\nu \rightarrow \infty$.
Eq.(\ref{51}) shows that
$\l^{(1)}_{OD} = 1$ in 1d. In fact from KW duality it is clear
that $\l^{(\nu)}_{OD} = 1$ for all $\nu$ in 1d.

Let us turn to the more interesting case of 2d, where
the partition function and other related
quantities are known exactly. In figure 3 we show
our  results for the energy.
The value of $\l^{(1)}_{OD}$ is $2.7239$ which is not far from the
critical value computed by other methods which is around $3.08$  \cite{PeEl}.

\subsection{First Excited State: the Gap}

A better determination of $\l_c$ will be for example to compute
the energy gap $\Delta$ in the spectrum of the Hamiltonian.
For values of $\l$ near and above $\l_c$ the singular behaviour
of $\Delta$ is characterized by an exponent $s$,

\begin{equation}
\Delta \sim ( \l - \l_c)^s , \;\; \l \geq \l_c
\label{52}
\end{equation}

\noindent
which is computed in the disordered region.

Assuming that the first excited state is translationally
invariant and that it belongs to the odd sector (R=-1)
, we propose the following ansatz,

\begin{equation}
\begin{array}{cl}
\p^{(1)}_{exc} =& \sum_{\m} \s^Z_{\m} \;
\e \left( \frac{\b}{2}  \sum_{\n,\mu}
\s^Z_{\n} \s^Z_{\n+ \mu} \right)
|0>
\end{array}
\label{52b}
\end{equation}

\noindent
where the variational parameter $\b$ can in principle be different
from the ground state parameter $\a$ appearing in (\ref{49}).
The norm of (\ref{52b}) is essentially given by the susceptibility
at zero magnetic field of a
classical Ising model (see appendix D). We have only
considered the 1d case where the result that we obtain coincides
quite surprisingly with the exact result,

\begin{eqnarray}
& E_{1}(\var ,\nu=1, d=1) - E_{0}(\var ,\nu=1, d=1)
& \nonumber \\
&=E_{1}(exact, d=1) - E_{0}(exact, d=1) = 2 (\l -1)&
\label{52bb}
\end{eqnarray}

For $d > 1$ the computation of $\Delta$
could be performed
using high-temperature expansions for the susceptibility.

\section{Concluding Remarks}

We have shown in the study of the ITF that a discrete versions
of generalized Jastrow wave functions are
very adequate to developp a perturbative/variational approach
to local lattice Hamiltonians. The information
encoded in the lowest orders of perturbation theory is amplified
to all orders by means  of a variational method. In this way
the ansatzs for the ground state of the hamiltonians
become less aprioristic, since they are subjected
to satisfy a set of strong constraints
which eliminates much of the arbitrariness.

An expected weakness of our method shows up for Hamiltonians
near criticality, as for the energy and
magnetization do not exhibit the desired singular
behaviour.
This can be understood from the fact that we
do not take into account all the scales involved near the
critical point. To make further progress one should fertilize
the P/V method with renormalization group ideas, in the spirit
of references \cite{DWY,PJP}.

{\bf Acknowledgements}

We would like to thank
M. Asorey, E. Brezin, J.G. Carmona, J. Ferrer,
C. G\'omez, M. Halpern, F. Jim\'enez,
M.A. Martin Delgado
and M.A.H. Vozmediano
for useful discussions.
This work has been supported the CICYT  grants  PB93-0302 (J.G.E.) and
PB92-1092 (G.S.).

\newpage

\section*{Appendix A:Variational Calculations in the Ordered Phase }

The basic matrix elements needed to compute the energy of the trial
wave function (\ref{39}) are the following,

\begin{equation}
\begin{array}{rl}
< \p^{(1)} | \p^{ (1) } > = & {\ch}^L  h \\
< \p^{(1)} | \sum_{\n} \s^X_{ \n } | \p^{(1)}> = &
L \; {\ch}^{L} h \;\;  \t \;  h \\
< \p^{(1)} | \sum_{\n} \s^Z_{ \n } | \p^{(1)}> = &
L \; {\ch}^{L-1} h  \\
< \p^{(1)}|\sum_{ \n,\mu} \s^Z_{\n} \;
\s^Z_{\n + \mu}  |  \p^{(1)}> = &  L d \; (\ch \; h )^{L-2}
\end{array}
\label{53}
\end{equation}

We have assumed that $h$ is real.
Using these equations the density energy of $\p^{(1)}$ becomes,

\begin{equation}
\en( \var ,\nu=1,d) = - \left( \l \; \t \; h + \frac{d}{{\ch}^2 h} \right)
\label{54}
\end{equation}

The minimum of (\ref{54}) is obtained for

\begin{equation}
\begin{array}{cl}
\l = 2 \; d \;\t h & {\rm if} \; \; \l \leq 2d \\
1 = \t h & {\rm if} \l \geq 2d
\end{array}
\label{55}
\end{equation}

Introducing (\ref{55}) into (\ref{54}) we obtain (\ref{40}).

The relevant matrix elements of the trial wave function
$\p^{(2)}$ are,

\begin{equation}
\begin{array}{rl}
< \p^{(2)}|\p^{(2)}> = & Z_L(\a,h) \\
< \p^{(2)}|\sum_{\n} \s^X_{\n} |   \p^{(2)}> = &
\frac{\partial}{ \partial h} Z_L(\a,h) \\
< \p^{(2)}|\sum_{\n} \s^Z_{\n} |\p^{(2)}> = &
L \; Z^{(site)}_{L-1}(\a,h) \\
< \p^{(2)}|\sum_{ \n,\mu} \s^Z_{\n} \;
\s^Z_{\n + \mu}  |  \p^{(2)}> = &  L\; d \;
\ch \a \;  Z^{(bond)}_{L-2} (\a,h)
\end{array}
\label{56}
\end{equation}

\noindent
where $Z_L(\a,h)$ is the partition function of a classical
statistical Ising model defined in a d-dimensional hypercubic
lattice,

\begin{equation}
Z_L(\a,h) = \frac{1}{2^L} \sum_{ \{ s_1, \cdots, s_L \} }
\e \left( \a \sum_{\n,\mu} \; s_{\n} s_{\n+\mu} +
h \sum_{\n} s_{\n}  \right)
\label{57}
\end{equation}

\noindent
where $s_{\n} = \pm$.

$Z^{(site)}_{L-1}(\a,h)$ is the same as $Z_L(\a,h)$
except for the removal of one
lattice variable located at
a single site, say {\bf L},
together with all possible couplings among this variable
and the ones in its neighbour,

\begin{equation}
Z^{(site)}_{L-1}(\a,h) = \frac{1}{2^{L-1}}
\sum_{ \{ s_1, \cdots, s_{L-1} \} }
\e \left( \a \sum_{\n,\n +\mu \neq {\bf L}} \; s_{\n } s_{\n+\mu} +
h \sum_{\n \neq {\bf L}} s_{\n} \right)
\label{58}
\end{equation}

Finally $Z^{(bond)}_{L-2}(\a,h)$ is the same as $Z_L(\a,h)$
except for the removal of
two lattice variables, say {\bf L} and {\bf L-1} forming
a single bond,
together with all possible couplings among this bond
and the remaining lattice variables, namely

\begin{equation}
Z^{(bond)}_{L-2}(\a,h) = \frac{1}{2^{L-2}} \sum_{ \{ s_1, \cdots, s_{L-2} \} }
\e \left( \a \sum_{\n,\n +\mu \neq {\bf L,L-1}} \; s_{\n } s_{\n+\mu} +
h \sum_{\n \neq {\bf L,L-1}} s_{\n} \right)
\label{59}
\end{equation}

To derive eqs.(\ref{56}) we  use the following
general formula which relates matrix elements of Pauli matrices
and statistical sums of Ising models,

\begin{equation}
<\uparrow| f( \s^X_1 , \s^X_2, \cdots, \s^X_M) |\uparrow>
= \frac{1}{2^M} \; \sum_{ s_1, \cdots, s_M}
f( s_1 , s_2, \cdots, s_M)
\label{60}
\end{equation}

\noindent
where $f(x_1, \cdots, x_M)$ is a generic function of $M$ variables.

Using eqs(\ref{56})
the energy of $\p^{(2)}$ becomes ,

\begin{equation}
E(\var ,\nu=2,d) = - \l \; \frac{\partial}{\partial h}
\ln Z_L(\a,h) - d \; L \; \frac{Z^{(bond)}_{L-2}(\a,h) }{Z_L(\a,h)}
\label{61}
\end{equation}

\noindent
and the magnetization reads,

\begin{equation}
< \s^Z > = <\p^{(2)}|  \s^Z_{\n}| \p^{(2)}> =
\frac{ Z^{(site)}_{L-1}(\a,h) }{ Z_L(\a,h)}
\label{62}
\end{equation}

In 1d these partition functions can be computed exactly in
the limit $L \rightarrow \infty$,

\begin{equation}
\begin{array}{rl}
Z^{(1d)}_L(\a,h) =&  z^L   \\
Z^{(1d, site)}_{L-1} (\a,h)=&
\frac{1}{2} \left[ \ch \; h +
\frac{ 1 +
e^{2 \a} \; \sh^2 h}{ \sqrt{ 1 + e^{4 \a} \; \sh^2 h} } \right]
z^{L-2} \\
Z^{(1d, bond)}_{L-2} (\a,h)=&
\frac{1}{2} \left[ \ch \; h +
\frac{ 1 + e^{2 \a}\; \sh^2 h}{\sqrt{ 1 + e^{4 \a} \; \sh^2 h }} \right]
z^{L-3} \\
\end{array}
\label{63}
\end{equation}

\noindent
where

\begin{equation}
z= \frac{1}{2} \; \left[ e^{\a} \ch \; h + e^{-\a}  \sqrt{ 1 +
e^{4 \a} \; \sh^2 h } \; \right]
\label{64}
\end{equation}

Introducing these equations into (\ref{61}) we obtain the energy
in 1d,

\begin{eqnarray}
&\en(\var ,\nu=2,d=1) = -2 \left(
\frac{  \l \; \sqrt{ s^2 -1  }}{2 \; s} + \frac{1}{s t}
- \frac{1 }{t^2} + \frac{ \sqrt{ 1 + t^2 - 2 s \; t} }{t^2} \right) &
\label{65} \\
& s = \sqrt{ 1 + e^{4 \; \a} \; \sh^2 h } & \label{66} \\
& t = s + \sqrt{ s^2 + e^{ 4 \; \a} - 1} & \label{67}
\end{eqnarray}

Minimization of (\ref{65}) with respect to $t$ and $s$ gives,

\begin{eqnarray}
&s = \frac{1}{9 t}  \left[ 3 t^2 + 2 + \sqrt{ 3 t^2 + 4 } \right]
& \label{68} \\
& \l = \frac{2 \sqrt{s^2 -1}} {t} \left( 1 +
\frac{ s^2 }{\sqrt{ 1 + t^2 - 2 s \; t}} \right) &
\label{69}
\end{eqnarray}

The magnetization in 1d is given, using the variables $s$ and
$t$, by

\begin{equation}
< \s^Z> = \frac{2}{t^2} \left[ t - \frac{1}{s} + \frac{1}{s}
\sqrt{ 1 + t^2 - 2 s \; t} \right]
\label{70}
\end{equation}

Plots of the energy (\ref{65}) and the magnetization (\ref{70}),
using (\ref{68}) and (\ref{69}), are given in figures 1 and 2 respectively.
Some numerical results are shown in table 1,
where we have included for comparison the results
obtained in reference \cite{QuWe} using a variational
renormalization group method. In this particular case the
variational energy obtained through the P/V method is better
than the one obtained using the RG method.
As for the magnetization, in
the RG method there is a finite value of $\l$ at which it
vanishes, whereas in the P/V method $\l$ has to go to $\infty$.

\begin{table}[h]
\begin{center}
\begin{tabular}{|c|c|c|c|c|}
\hline
$\lambda$ & energy (var)& energy (RG) & $< \sigma^Z >$(var)&
$< \sigma^Z >$(RG) \\ \hline
0.2 & - 1.010 024 99 [525]&-1.010 000  & 0.994 912 [0]& 0.994 93 \\ \hline
0.4 & - 1.040 399    [417]&-1.040 306  & 0.978 56 [44]& 0.978 83 \\ \hline
0.9 & - 1.212 4      [60]&-1.210 8  & 0.859 3 [125]& 0.860 4 \\ \hline
1.0 & - 1.264 9      [732]&-1.262 9 & 0.8 [0]& non zero \\ \hline
\end{tabular}
\caption{ Energy density and magnetization
of the 1d ITF
obtained by the
$\nu = 2$ variational method and the RG methods of reference {\protect
{\cite{QuWe}}}.
 Numbers in brackets  are last digits of the exact results. }
\end{center}
\label{70tb}
\end{table}

\section*{Appendix B: Variational Calculations in the Disordered Phase }

The matrix elements for the ansatz (\ref{49}) are,

\begin{equation}
\begin{array}{rl}
< \p^{(1)}|\p^{(1)}> = & Z_L(\a) \\
< \p^{(1)}|\sum_{\n} \s^X_{\n} |   \p^{(1)}> = &
L \; Z^{(site)}_{L-1}(\a) \\
< \p^{(1)}|\sum_{ \n,\mu} \s^Z_{\n} \;
\s^Z_{\n + \mu}  |  \p^{(1)}> = &
\frac{\partial}{ \partial \a} Z_L(\a) \\
\end{array}
\label{71}
\end{equation}

\noindent
where $Z_L(\a)= Z_L(\a,h=0)$ and
$Z^{(site)}_{L-1}(\a)= Z^{(site)}_{L-1}(\a,h=0)$.

The energy reads,

\begin{equation}
E(\var , \nu=1, d) =
- \frac{\partial}{ \partial \a} \ln Z_L(\a)
- \l \; L \;
\frac{ Z^{(site)}_{L-1}(\a) }{ Z_{L}(\a) }
\label{72}
\end{equation}

In 1d one has,

\begin{eqnarray}
& Z^{(1d)}_{L}(\a) = \ch^L \; \a & \label{73} \\
& Z^{(1d,site)}_{L-1}(\a) = \ch^{L-2} \; \a & \label{74}
\end{eqnarray}

which yields the energy density,

\begin{equation}
e( \var , \nu=1, d=1) = - \left( \t \; \a + \frac{ \l}{
\ch^2 \a} \right)
\label{75}
\end{equation}

The minimum of this expresion is obtained for,

\begin{equation}
\begin{array}{cl}
\frac{1}{\l} = 2 \; \t \; \a & {\rm if} \; \l \geq 1/2 \\
1 = \t \; \a &  {\rm if} \; \l \leq 1/2 \end{array}
\label{76}
\end{equation}

Substituting (\ref{76}) into (\ref{75}) one obtains eq.(\ref{50}).

To study the 2d disordered case we need the well known
results obtained by Onsager in his study on the
2d Ising model \cite{On}.
Calling $f_{2d}= - \lim_{L \rightarrow \infty} \frac{1}{L} \log Z_L(\a)$
the free energy per site, $u_{2d} =
\partial f_{2d}/\partial \a $ the internal energy
and $c_{2d}= - \a^2 \partial u_{2d}/\partial \a $
the specific heat we have,

\begin{eqnarray}
&f_{2d} = - \ln \cosh(2 \a) \; - \frac{1}{\pi} \;
\int^{\pi/2}_{0} \; d \phi \ln \frac{ 1+
\sqrt{ 1 - q^2 \; \sin^2 \phi} }{2} & \nonumber \\
& & \nonumber \\
&u_{2d} = - \coth(2 \a) \left\{
1 + \frac{2}{\pi} q'
\; K(q) \right\} & \label{77} \\
& & \nonumber \\
& c_{2d} = \frac{4 \a^2}{\pi} \coth^2(2 \a) \;
\left\{ K(q) - E(q) \right. & \nonumber \\
& \left. + \frac{1}{2} (q'-1)
\left( \frac{ \pi}{2} + q
K(q) \right) \right\} & \nonumber
\end{eqnarray}

\noindent
where

\begin{eqnarray}
&q= \frac{ 2 \sinh 2 \a}{ \cosh^2( 2 \a)}& \nonumber \\
&q'= 2 \; \t^2( 2 \a) -1 &
\label{78}
\end{eqnarray}

and $K(q)$ and $E(q)$ are the elliptic integrals,

\begin{equation}
K(q) = \int^{\pi/2}_0 \frac{ d \phi}{
\sqrt{ 1 - q^2 \; \sin^2 \phi} } \;\;, \;\;
E(q) = \int^{\pi/2}_0 d \phi
\sqrt{ 1 - q^2 \; \sin^2 \phi}
\label{79}
\end{equation}

The computation of $Z^{(2d,site)}_{L-1}$
is a bit more complicated but it can be related
to the original partition function through
two point correlation functions as follows,

\begin{eqnarray}
&\frac{Z^{(2d,site)}_{L-1}}{ Z^{(2d)}_{L} } = \frac{1}{2}
\left[ 2 + x^2 - x ( 5 + x^2) < s_0 s_{\mu_1}> +
x^2 ( <  s_0 s_{2 \mu_1} > + 2 < s_0 s_{\mu_1 + \mu_2} > ) \right]
& \label{80}
\end{eqnarray}

\noindent
where $x= \t \a$ and
$s_0,s_{\mu_1},s_{2 \mu_1},s_{\mu_1 + \mu_2}$ are classical Ising
variables located at the points $0, \mu_1,2 \mu_1, \mu_1 + \mu_2 $
respectively.
The correlators entering eq.(\ref{80}) has been computed in
reference \cite{Fi},

\begin{equation}
\begin{array}{rl}
<  s_0 s_{ \mu_1} > = &
\frac{1}{2} {\rm cotanh}(2  \a) \left[ 1 + \frac{2}{\pi}
q' K(q) \right] \\
< s_0 s_{\mu_1 + \mu_2} > = &
\frac{1}{\pi} {\rm cotanh}^2(2  \a) \left[ E(q) + q' K(q)
\right] \\
<  s_0 s_{2 \mu_1} > = &
\frac{1}{2} {\rm cotanh}^2(2  \a) - \left(\frac{2}{\pi q}\right)^2
\left[ E^2(q) - 2 q' K(q) E(q) + (q')^3 K^2(q) \right]
\end{array}
\label{81}
\end{equation}

Introducing equations (\ref{77}),(\ref{80}) and (\ref{81})
into (\ref{72}) and minimizing this energy numerically
we obtain figure 3, where we have also include the mean field result
(\ref{40}) for 2d
and the lower bound given by eq.(\ref{28b}).

\section*{Appendix C: Other P/V solutions}

We mentioned in section 2 the non uniqueness of the solution of the
perturbative equations (\ref{10}). This in turn lead us to look
for solutions which in the limit $\l \rightarrow \infty$ will flow
to the ground state of $H_1$. This is the reason to impose
conditions (\ref{23}). To get further motivation for the need of
these conditions it is illustrative to investigate
another solutions to equations (\ref{10}) than those given in section
3. This will be done only at the lowest order $\nu =1$.

In the ordered phase an antihermitean solution of eq.(\ref{10}) is given
by,

\begin{equation}
U_1 = \frac{ - \i }{ 4 d} \; \sum_{\n} \s^Y_{\n}
\label{82}
\end{equation}

\noindent
which leads to the following ansatz,

\begin{equation}
\begin{array}{cl}
\p^{(1)}& = \e \left(- \frac{\i \theta}{2} \sum_{\n} \s^Y_{\n}
\right) |\uparrow> \\
& =  \prod_{\n} \left( {\rm cos} \frac{ \theta}{2} \; | \uparrow>+
{\rm sin} \frac{ \theta}{2} \; | \downarrow> \right) \end{array}
\label{83}
\end{equation}

\noindent
where $\theta$ is a real parameter.
The variational energy of this state can be computed as in
appendix A obtaining,

\begin{equation}
\en({\rm var}, \nu=1,d) = - \left( \l \; {\rm sin} \theta +
d \;  {\rm cos}^2 \theta \right)
\label{84}
\end{equation}

Whose mimimun is achieved at $\theta$ given by,

\begin{equation}
{\rm sin} \theta_0 = \left\{
\begin{array}{cl}
\frac{\l}{2 d} & \l \leq 2 d \\
1 & \l \geq 2d \\
\end{array} \right.
\label{85}
\end{equation}

Substituting this result in (\ref{84}) we get the same energy
as the one given by eq.(\ref{40}), which was obtained with the
hermitean solution (\ref{39}). Indeed if we compare eqs.(\ref{39})
and (\ref{83}) we see that the trigonometric variable $\theta$ can
be related to the hyperbolic variable $h$ by means of
the Gudermannian function,

\begin{equation}
\theta = gd(h) = 2 {\rm arctanh}(\en^h) - \frac{\pi}{2}
\label{86}
\end{equation}

The mean field value computation of reference \cite{dG}
is performed in terms of the variable
$\theta$ whose physical meaning is the
semiclassical rotation  suffered by
the spin under the action of the external field $\l$,

\begin{equation}
< \s^Z > = {\rm cos} \theta
\label{87}
\end{equation}

Let us move on to the disordered case.
We shall make the following choice of the operator
$U_1$,

\begin{eqnarray}
&U_1 = \epsilon \; \sum_{\n,\mu} \left( \vec{ {\rm r}}
\cdot \vec{\s_{\n}} \right) \;
\left( \vec{ {\rm r}} \cdot \vec{ \s_{\n+ \mu}} \right)&
\nonumber \\
&U_1^\dagger = \epsilon^2 \; U_1&
\label{88}
\end{eqnarray}

\noindent
where $\vec{ {\rm r}}= (0,y,z)$
is a vector with real components
and $\epsilon^4 =1$.
There are 4 possible solutions of the first order perturbative
equation (\ref{10}) which we display in table
2 together with the corresponding
variational energies in the 1d case.

\begin{table}[h]
\begin{center}
\begin{tabular}{|c|c|c|c|}
\hline
Solutions & $\epsilon$ & $\vec{ {\rm r}}$ & $\en(var,\nu=1,d=1) $ \\ \hline
I &  1 &$ (0,0,\frac{1}{2})$ &$
\left\{  \begin{array}{ll}
- (\l + \frac{1}{4 \l}) & \l \geq \frac{1}{2} \\
-1 & \l \leq \frac{1}{2}
\end{array} \right. $
\\ \hline
II &$ -1$ & $(0,\frac{1}{2},0) $   &
$- \frac{2}{27} \left[ \l \; (9- \l^2) + (\l^2 + 3)^{3/2} \right]
$ \\ \hline
III & $\i$ &$ (0, \frac{1}{2 \sqrt{2}}, \frac{1}{2 \sqrt{2}})
$ &$ -\frac{1}{2} \left[ \l + \sqrt{ 1 + \l^2} \right] $  \\ \hline
IV &$ - \i$ &$ (0,- \frac{1}{2 \sqrt{2}}, \frac{1}{2 \sqrt{2}})$
& $ -\frac{1}{2} \left[ \l + \sqrt{ 1 + \l^2} \right]    $
\\ \hline
\end{tabular}
\caption{Variational states and  energies for the $\nu =1, d=1$
disordered case  }
\end{center}
\label{88tb}
\end{table}

We observe that only solution I, which is the one used
in section 3, goes in the limit $\l
\rightarrow 0$ to the correct value of the energy of the ordered
region. Indeed this solution satisfies conditions (\ref{23})
while the others do not. Table 2 shows that for the ansatz (\ref{88})
there exists no equivalence between hermitean and antihermitean
solutions as it happens in the mean field case.

\section*{Appendix D:Calculation of the Mass Gap}

The relevant matrix elements for this state are,

\begin{equation}
\begin{array}{rl}
< \p^{(1)}_{exc}|\p^{(1)}_{exc}> = & \tilde{Z}_L(\b) \\
< \p^{(1)}_{exc}|\sum_{\n} \s^X_{\n} | \p^{(1)}_{exc}> = &
L \left( \tilde{Z}^{(site)}_{L-1}(\b) - Z^{(site)}_{L-1}(\b) \right) \\
< \p^{(1)}_{exc}|\sum_{ \n,\mu} \s^Z_{\n} \;
\s^Z_{\n + \mu}  |  \p^{(1)}_{exc}> = &
\frac{\partial}{\partial \b} \; \tilde{Z}_L(\b)
\end{array}
\label{89}
\end{equation}

\noindent
where

\begin{equation}
\tilde{Z}_L(\b) =
\frac{\partial^2}{\partial h^2} \; {Z}_L(\b,h)|_{h=0}
\label{90}
\end{equation}

\noindent
and similarly for $\tilde{Z}^{(site)}_L(\b)$.
The expression of the energy of the state (\ref{52b}) is,

\begin{equation}
E_{1}(\var, \nu=1,d) = -\left(
\frac{\partial}{\partial \b} \;\ln  \tilde{Z}_L(\b)
+ \frac{L \l}{\tilde{Z}_L(\b) } \left(
\tilde{Z}^{(site)}_{L-1} - Z^{(site)}_{L-1} \right) \right)
\label{91}
\end{equation}

A long but straightforward calculation
using eqs(\ref{63}) yields,

\begin{equation}
E_{1}(\var , \nu=1, d=1) = -L \; \left(
\t \; \b + \frac{\l}{\ch^2 \b} \right) + 2 (\l -1)
\label{92}
\end{equation}

\noindent
whose minimization gives (\ref{52bb}).
The value of $\beta$ coincides with that of $\a$ given by
eq.(\ref{76}).

\newpage

\section*{Figure Captions}

Fig.1.- Energy versus $\l$ in 1d:
exact (continuous line),  mean field (+++) and
$\nu=2$ variational in the ordered regime ($\cdots$).

\noindent
Fig.2.- Magnetization $<\s^Z>$ versus $\l$ in 1d:
exact (continuous line),  mean field (+++)  and
$\nu=2$ variational in the ordered regime ($\cdots$).

\noindent
Fig.3.- Energy versus $\l$ in 2d:
lower bound (\ref{28b}) (continuous line),
$\nu =1$ ordered phase (+++) and
$\nu=1$ disordered phase ($\cdots$).
\newpage
%%%%%%%%%%%%%%%%%%%%%%%%%%%%%%%%%%%%%%%%%%%%%%%%%%%%%%%%%%%%%%%%
%%%%%%%%%%%%%%%%%%%%%%%%%%%%%%%%%%%%%%%%%%%%%%%%%%%%%%%%%%%%%%%%
%BEGIN FIGURES
\bigskip
% BEGIN FIG1
\epsfbox[80 350 212 692]{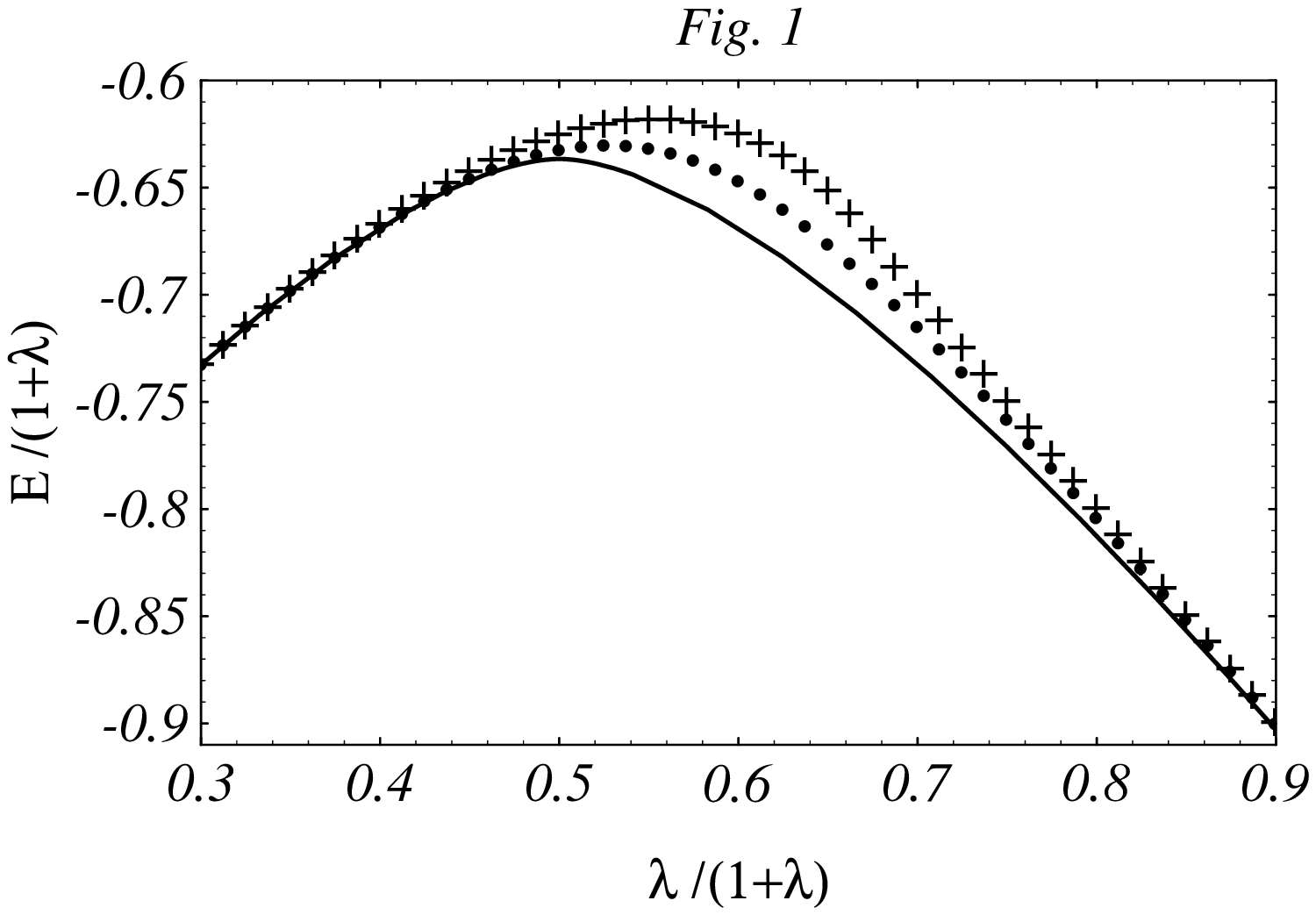}
% END FIG1
\vfill\eject
\bigskip
% BEGIN FIG2
\epsfbox[80 350 212 692]{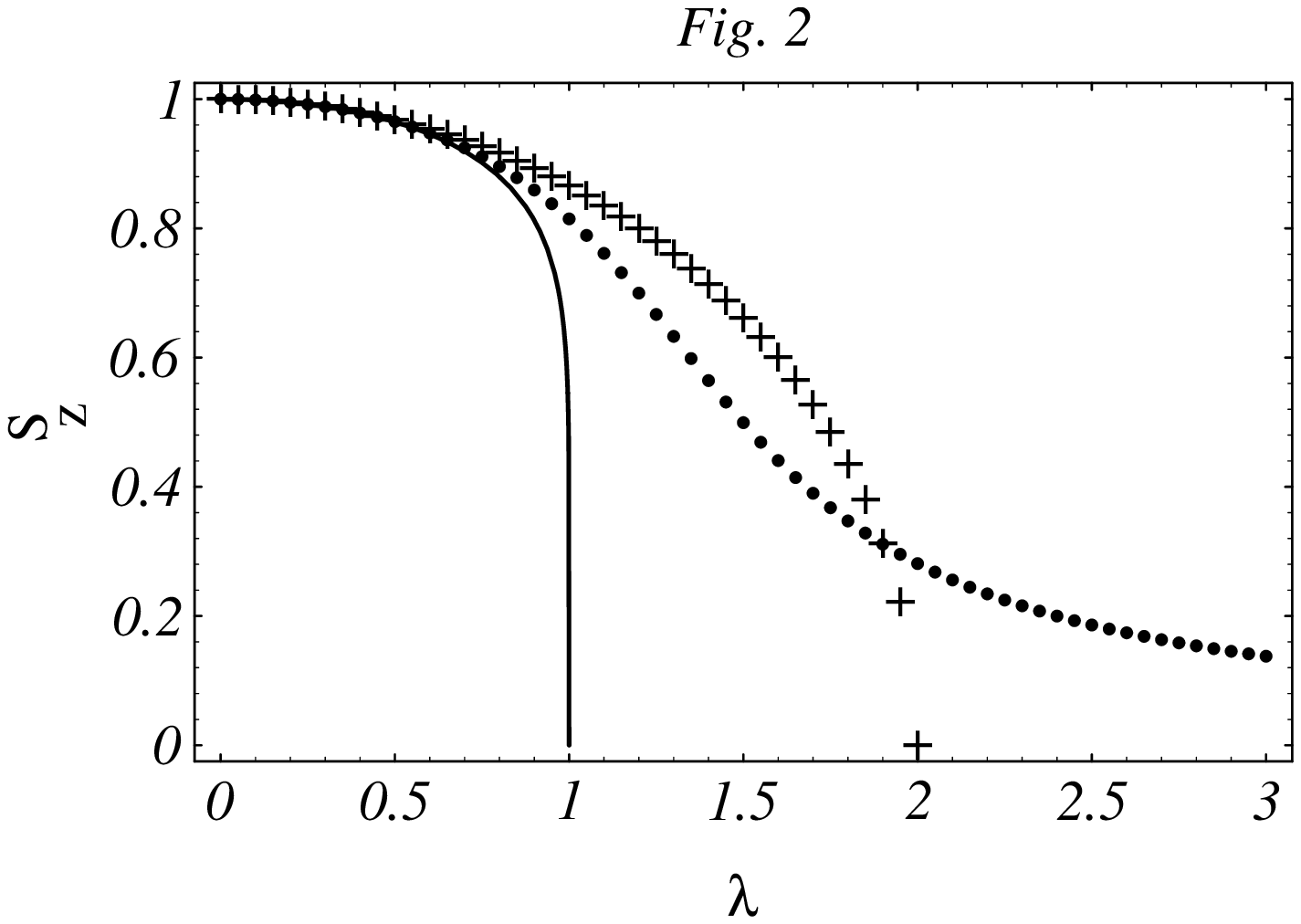}
% END FIG2
\vfill\eject
\bigskip
% BEGIN FIG3
\epsfbox[80 350 212 692]{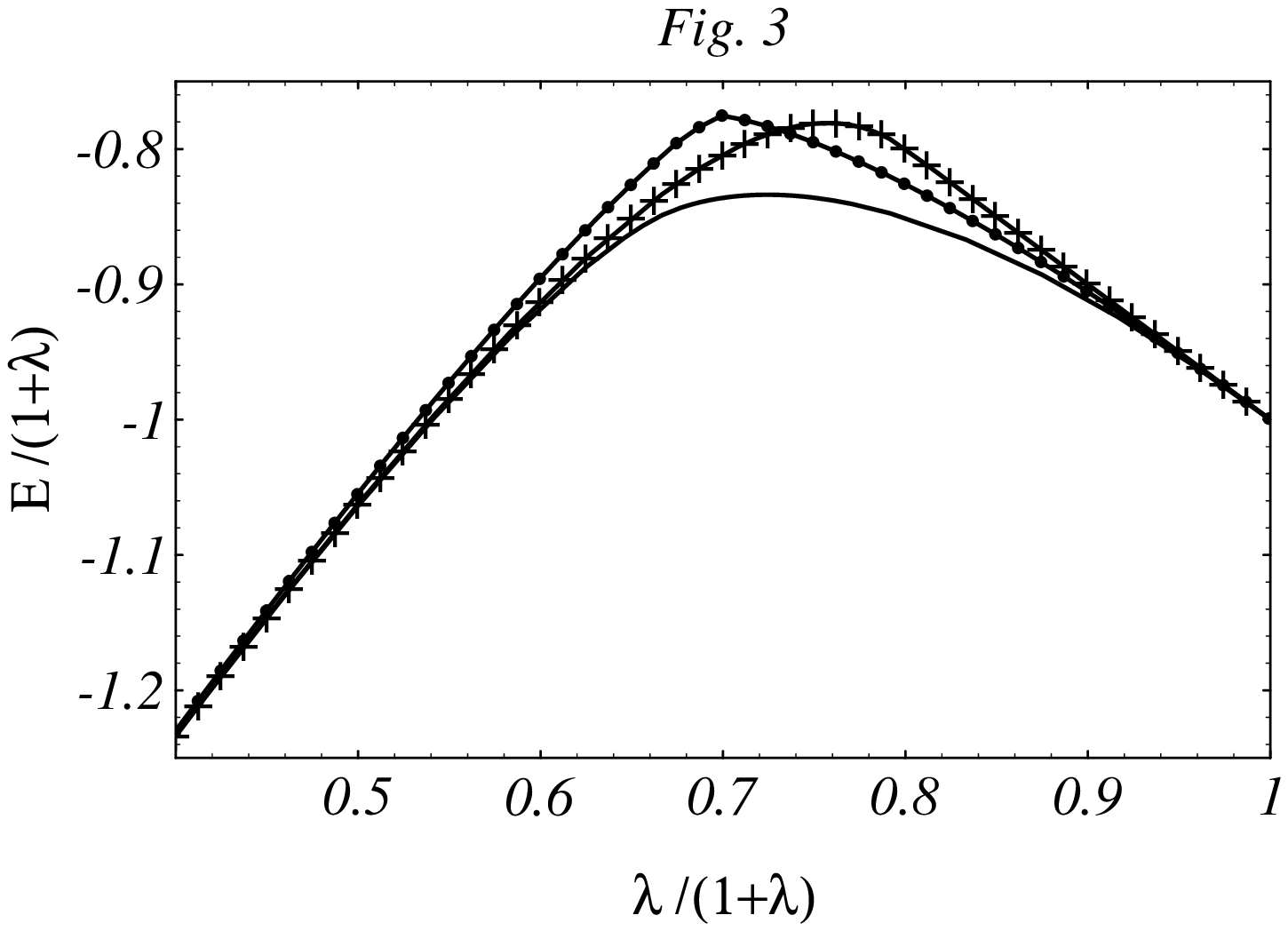}
% END FIG3
% END FIGURES
%%%%%%%%%%%%%%%%%%%%%%%%%%%%%%%%%%%%%%%%%%%%%%%%%%%%%%%%%%%%%%%%
%%%%%%%%%%%%%%%%%%%%%%%%%%%%%%%%%%%%%%%%%%%%%%%%%%%%%%%%%%%%%%%%

\end{document}